\documentclass[pra,twocolumn,aps,superscriptaddress,showpacs]{revtex4}

\usepackage{times}
\usepackage{graphicx} 
\usepackage{dcolumn} 
\usepackage{bm} 

\begin{document}


\title{Saturation of Cs$_2$ Photoassociation in an Optical Dipole Trap}

\author{S. D. Kraft}
\author{M. Mudrich
  \footnote{present address: Laboratoire Aim{\'e} Cotton, CNRS, 
  B{\^a}timent 505, Campus d'Orsay, 91405 Orsay, France}
}
\author{M. U. Staudt
  \footnote{present addres: University of Geneva,
  20 rue de l'École-de-Médecine, 1211 Geneva 4, Switzerland}
}
\author{J. Lange}
\affiliation{Physikalisches Institut, Universit{\"a}t Freiburg,
  Hermann-Herder-Stra{\ss}e 3, 79104 Freiburg, Germany}
\author{O. Dulieu}
\affiliation{Laboratoire Aim{\'e} Cotton, CNRS, B{\^a}timent 505, Campus 
  d'Orsay, 91405 Orsay, France}
\author{R. Wester}
\author{M. Weidem{\"u}ller}
\affiliation{Physikalisches Institut, Universit{\"a}t Freiburg,
  Hermann-Herder-Stra{\ss}e 3, 79104 Freiburg, Germany}

\date{\today}

\begin{abstract}
We present studies of strong coupling in single-photon photoassociation of
cesium dimers using an optical dipole trap. A thermodynamic model of the trap
depletion dynamics is employed to extract absolute rate coefficents. From the
dependence of the rate coefficient on the photoassociation laser intensity, we
observe saturation of the photoassociation scattering probability at the
unitarity limit in quantitative agreement with the theoretical model by Bohn
and Julienne \cite{bohn1999:pra}. Also the corresponding power broadening of
the resonance width is measured. We could not observe an intensity dependent
light shift in contrast to findings for lithium and rubidium, which is
attributed to the absence of a $p$ or $d$-wave shape resonance in cesium.
\end{abstract}


\maketitle



Ultracold thermal and quantum degenerate atomic ensembles have allowed to
investigate powerful coupling schemes between continuum scattering states of
two atoms and bound molecular states of the corresponding dimer. Both
photoassociation light \cite{miller1993:prl,fioretti1998:prl} or magnetic
field sweeps across Feshbach resonances
\cite{regal2003:nat,herbig2003:sci,strecker2003:prl} have lead to the
formation of ultracold molecules. Three-body recombination of fermionic atoms
near a Feshbach resonance has finally lead to the creation of molecular Bose
Einstein condensates
\cite{jochim2003:sci,greiner2003:nat,zwierlein2003:prl}. Recently, molecular
photoassociation has been performed using a two-photon Raman process in a
two-atom Mott insulator phase \cite{rom2004:prl}, which may prove to be a
route to a molecular Mott insulator and, subsequently, an alternative way to a
molecular BEC \cite{jaksch2002:prl}. In contrast to this, the successful
formation of ultacold molecular ensembles in the absolute rovibrational ground
state, which is important if one wants to study molecular collisions and
reactions over long interaction times, is still at large. Future routes for
the production of absolute ground state molecules from a pair of atoms will
certainly involve Raman type coupling schemes
\cite{lisdat2002:ejd,araujo2003:jcp,schloder2003:pra}. A very interesting
approach to transfer highly vibrationally excited molecules into the
vibrational ground state proposes the application of optimally controlled
femtosecond laser pulses \cite{koch2004:pra}.

In order to understand the regime of strong coupling in photoassociation,
which is important for both continuous and femtosecond Raman processes, we
report in this Rapid Communication investigations of saturation effects in the
first photoassociation step from the two-atom continuum to an excited
molecular level. This is also the determining step in the overall molecule
formation rate \cite{dion2001:prl}. Using an ultracold ensemble of cesium
atoms trapped in a quasi-electrostatic optical trap \cite{grimm2000:adv} we
are able to employ trap loss measurements to extract absolute photoassociation
rate coefficients \cite{wester2004:apb}, from which the dependence of the rate
coefficient on the photoassociation laser intensity is derived.

Previously, Schl{\"o}der {\it et al.}  have investigated strong coupling in a
$^6$Li$^7$Li two-species magneto-optical trap and observed broadening and
center shifts of several photoassociation resonances
\cite{schloder2002:pra}. Comparing their results to the semiclassical model of
Bohn and Julienne for strong-field photoassociation \cite{bohn1999:pra} a
saturation intensity of 30-50\,W/cm$^2$ was derived \cite{schloder2002:pra}.
McKenzie {\it et al.}  \cite{mckenzie2002:prl} and Prodan {\it et al.}
\cite{prodan2003:prl} have studied photoassociation out of a Bose-Einstein
condensate of Na and $^7$Li, respectively, to determine the limit for the
photoassociation rate coefficient. While McKenzie {\it et al.} have  observed a
linear relationship between laser intensity and association rate up to about
1\,kW/cm$^2$, Prodan {\it et al.} clearly find that the photoassociation rate
saturates at the unitarity limit of the inelastic scattering cross section,
due to probability conservation in the quantum mechanical scattering process
\cite{prodan2003:prl}.  These results are conveniently interpreted in the
framework of the two-body scattering model of Ref.\ \cite{bohn1999:pra}, as
long as the low intensity regime is concerned. Effects of high
photoassociation laser intensities have been investigated in several other
papers, emphasizing the multichannel character of the photoassociation process
which have to be described within a molecular dressed-state approach. Specific
discussions on cold sodium and rubidium photoassociation are proposed in Ref.\
\cite{simoni2002:pra}, on strontium photoassociation in Ref.\
\cite{montalvao2001:pra}, and on photoassociation in a Na Bose-Einstein
condensate in Ref.\ \cite{gasenzer2004:pra}.



The photoassociation experiments are performed with a thermal ensemble of
cesium atoms trapped in an optical dipole trap formed by the focus of a CO$_2$
laser \cite{wester2004:apb}. As described in Refs.\
\cite{engler2000:pra,mosk2001:apb}, the cesium atoms are loaded from a
magneto-optical trap which is superimposed in the focus of the CO$_2$
laser. The MOT is operated in five-beam configuration and loads from a Zeeman
slowed beam up to $10^8$ particles at a density of $10^9\,\mbox{cm}^{-3}$ as
inferred from absorption imaging.

After turning of the magnetic field of the MOT a brief molasses cooling phase
transfers about $5\times10^5$ atoms at a density of
$5\times10^{11}\,\mbox{cm}^{-3}$ and a temperature of 40\,$\mu$K into the
optical dipole trap \cite{engler2000:pra,mosk2001:apb}.  For the cold cesium
atoms the CO$_2$ laser focus with its potential depth of 0.8\,mK represents a
harmonic trap with axial and radial trap frequencies of 12.8\,Hz and 625\,Hz
respectively. The lifetime of the atoms in the dipole trap is of the order of
100\,s, due to collisions with residual gas atoms.

The photoassociation light is provided by a widely tunable Titanium:Sapphire
laser (Coherent MBR 110) system with a typical output power of 200\,mW, and a
line width of about 100\,kHz. Relative frequency changes are monitored by
measuring the transmission signal of the Ti:Sapphire laser through a confocal
cavity with a free spectral range of 500\,MHz. In order to decrease the
frequency spacing of the transmission peaks, two side-bands are modulated onto
the Ti:Sapphire beam at $\pm$166\,MHz using a double-pass AOM setup. The
length of the cavity is stabilized by locking one of the cavity mirrors with a
piezo actuator to the fringe of the transmission signal of a superimposed
spectroscopy stabilized diode laser. This provides a relative frequency
accuracy of about 5\,MHz. The absolute laser frequency is measured with a
commercial wavemeter (Burleigh WA 1000) with an accuracy of 500\,MHz. The
photoassociation laser is passed through the trapped cesium cloud in the focus
of the CO$_2$ laser at an angle of $22.5^\circ$ with respect to the CO$_2$
laser beam. The width of the Ti:Sapphire beam at the trap center amounts to
about 150\,$\mu$m. The intensity of the photoassociaton beam has been changed
between 30 to 400\,W/cm$^2$.

Once the cesium atoms are loaded into the dipole trap, the shutter of the
photoassociation laser is opened and the atom cloud is illuminated for
1\,s. Then the CO$_2$ laser light is extinguished and all remaining cesium
atoms are recaptured into the magneto-optical trap. The number of recaptured
cesium atoms is obtained from the fluoresence signal of the MOT with an
accuracy of 30\,\%. Since the storage time of atoms inside the dipole trap can
range up to 100\,s the atom loss signal is a direct measure for the formation
of cold molecules through photoassociation.



\begin{figure}[ht]
\includegraphics[width=8cm]{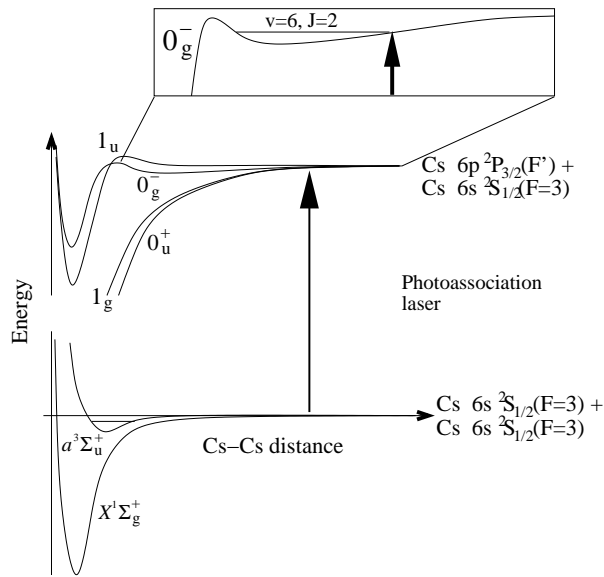}
\caption{\label{fig:photoassociation} Overview of the important potential
  energy curves for the photoassociation of cesium molecules.}
\end{figure}

\begin{figure}[ht]
\includegraphics[width=8cm]{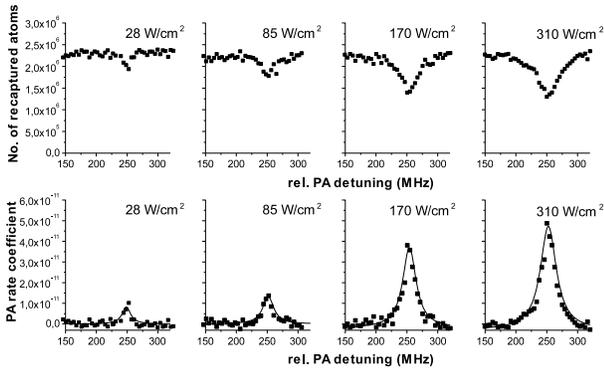}
\caption{\label{fig:saturated_spectra} Upper panels: trap loss spectra of the
studied photoassociation resonance $v=6,J=2$ of the 0$_g^-$ outer well for
different laser intensities. Lower panels: Spectra of the photoassociation
rate coefficients extracted from the trap loss signal. The solid lines
represent Lorentzian fits to the resonances.}
\end{figure}

To investigate how and at which intensities the photoassociation scattering
rate saturates we have measured the dependence of the association rate into
the $v=6$, $J=2$ rovibrational states of the 0$_g^-$ outer potential well (see
Fig.\ \ref{fig:photoassociation}) on the photoassociation laser intensity. The
atom loss spectrum of this photoassociation resonance is shown for four
different laser intensities in the upper panels of Fig.\
\ref{fig:saturated_spectra}. It can be seen that the width increases with
intensity, whereas the increase in the number of lost atoms due to
photoassociation slows down for large intensities. The peak position on the
other hand remains unchanged. In order to extract absolute photoassociation
rate coefficients from the atom loss spectra we have developed an analyic
thermodynamic model of the trap depletion dynamics that takes into account the
decrease in pair density and increase in cesium temperature over the course of
the photoassociation process \cite{wester2004:apb}. Using this model the atom
loss spectra are converted into rate coefficients as shown in the lower panels
of Fig.\ \ref{fig:saturated_spectra}.  The overall accuracy of the rate
coefficient values is estimated to be 40\,\% due to the accuracy in particle
number and temperature determination \cite{wester2004:apb}. The rate
coefficients are found to exhibit a stonger increase with laser intensity as
the atom loss spectra, which is a consequence of the saturation due to trap
dynamics that is unfolded in the rate coefficient spectra. Nevertheless, both
a saturation effect and a broadening of the rate coefficient spectra in the
lower panel of Fig.\ \ref{fig:saturated_spectra} are observed. These are
attributed to intrinsic saturation of the single-photon photoassociation
process.

\begin{figure}[ht]
\includegraphics[width=8cm]{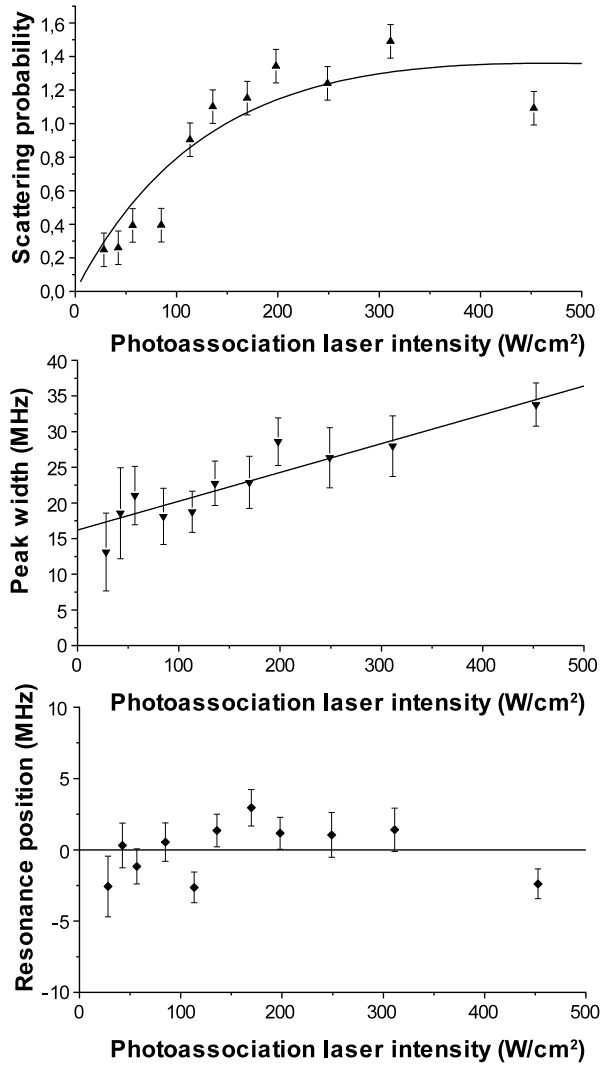}
\caption{\label{fig:width_rate} Photoassociation scattering probability (upper
  panel), resonance width (middle panel) and resonance line shift (lower
  panel) as a function of the photoassociation laser intensity. The solid
  lines represent the fit to the saturation model of Bohn and Julienne.}
\end{figure}

By fitting a single Lorentzian to the $v=6,J=2$ resonances, as shown in Fig.\
\ref{fig:saturated_spectra}, we obtain the maximum photoassociation rate
coefficient, the resonance width and the relative resonance position as a
function of the photoassociation laser intensity (see Fig.\
\ref{fig:width_rate}). The peak width is corrected for the 3.5\,MHz
inhomogeneous AC stark shift that the atoms experience in the dipole trap. The
thermal broadening of the resonances at the temperature of the cesium atoms of
40\,$\mu$K corresponds to less than 1 MHz and is also corrected for.

From the fitted rate coefficient $G$ the quantum mechanical scattering
probability $p_{\mbox{scatt}}$ is given by
\begin{equation}
p_{\mbox{scatt}} = \frac{G}{G_{\mbox{unitarity}}}
\end{equation}
where the rate coefficient at the unitarity limit denotes
\begin{equation}
G_{\mbox{unitarity}} = \langle \frac{\pi}{k^2} v_r \rangle_{v_r}
= \hbar^2
\sqrt{ \frac{2 \pi}{\mu^3 k_{\mbox{B}} T} }.
\end{equation}
When plotting the scattering probability as a function of the intensity for
intensities up to 500\,W/cm$^2$ (see Fig.\ \ref{fig:width_rate}a) saturation
is clearly observable. Within the experimental accuracy the measured
scattering probability always stays below unity, in agreement with the
unitarity limit. The resonance width increases linearly with intensity (see
Fig.\ \ref{fig:width_rate}b), whereas the position of the resonance remains
unchanged (see Fig.\ \ref{fig:width_rate}c).

These results are compared to the theoretical model of Bohn and Julienne for
photoassociation in strong laser fields \cite{bohn1999:pra}, giving the
scattering probability:
\begin{equation}
\label{eq:sat1}
p_{\mbox{scatt}} = \frac{\gamma \Gamma}{\frac{1}{4} (\gamma + \Gamma)^2}, =
\frac{I I_S}{\frac{1}{4} (I + I_S)^2},
\end{equation}
where $\gamma$ is the natural linewidth of the transition. $\Gamma$ denotes
the coupling strength defined by the overlap of the initial continuum
wavefunction $| \phi (E_{\rm kin} \rightarrow 0) \rangle$ with the final bound
state wavefunction $|\psi(v)\rangle$
\begin{equation}
\Gamma = 2 \pi (V_{eg})^2
| \langle \psi(v) | \phi (E_{\rm kin} \rightarrow 0) \rangle |^2.
\end{equation}
$V_{eg}$ is the Rabi frequency of the electronic transition, which is
proportional to the square-root of the laser intensity. The saturation
intensity $I_S$ is introduced according to $I / I_S = \Gamma / \gamma$. In
this formalism the resonance width is given by
\begin{equation}
\label{eq:sat2}
\delta f_{\mbox{FWHM}} = \gamma + \Gamma = \gamma ( 1 + \frac{I}{I_S} ) .
\end{equation}

The saturation intensity is obtained from the measured data for the scattering
probability and the resonance width by a fit with Eqs.\ \ref{eq:sat1} and
\ref{eq:sat2} (see Fig.\ \ref{fig:width_rate}). The two fits agree very well
with each other and yield a saturation intensity of 460$\pm$90 and
450$\pm$150\,W/cm$^2$, respectively. The fit to the scattering probability
used the maximum scattering probability as a second free parameter because the
overall accuracy of the scattering probabilities of about 40\,\%, due to the
accuracy of the extracted rate coefficients, is larger than the accuracy of
the fit. The fit value of $1.36\pm0.07$ for the maximum scattering probability
is in good agreement with unity within this experimental accuracy.

By extrapolation to zero intensity the natural line width $\gamma$ of the
photoassociation transition can be extracted. From the fit, a value of
$17\pm2$\,MHz is derived.  This value is larger than the predicted theoretical
resonance width of 10.4\,MHz that amounts to twice the atomic line width of
the $6s-6p$ transition.  Thus, the measured width may be the consequence of
additional broadening mechanisms.  One mechanism that could be responsible for
the discrepancy between the prediction and the experiment is the lifting of
the degeneracy of the magnetic sub-levels of the $J=2$ state in the electric
field of the CO$_2$ laser \cite{rost1992:prl}. We have estimated the energy
difference between the different $m_J$ levels to be about 7\,MHz, assuming
that the molecular polarizability along the internuclear axis is twice the
atomic polarizability. This interpretation of aligned molecules in the field
of the CO$_2$ laser trap will be investigated further in future experiments.
Another possible explanation could be found in Ref.\
\cite{pellegrini2003:phd}, where photoassociation is described within the
molecular dressed state approach at low intensities. Due to the very low
kinetic energy in the entrance channel, the continuum wavefunction and the
photoassociation wavefunction are predicted to be coupled at low
photoassociation intensities (typically 10\,W/cm$^2$) to the last bound state
lying very close to the Cs($6s$)+Cs($6s$) limit, inferred by its large
scattering length. This contribution is not included into the model of Ref.\
\cite{bohn1999:pra}. The consequent multichannel character of the
photoassociation process described now as a two-level plus continuum coupled
system, would induce a non-linear variation of the linewidth, which could well
converge to the expected doubled atomic value. This hypothesis, which is also
discussed in Ref.\ \cite{montalvao2001:pra}, is not easy to check in
experiments due to the weakness of the photoassociation signal at very low
intensities.

Ref.\ \cite{bohn1999:pra} also predicts a linear shift of the absolute
resonance position, previously observed in a $^7$Li quantum degenerate gas
\cite{prodan2003:prl}, and in a Na Bose-Einstein condensate
\cite{mckenzie2002:prl}. In the latter case, the shift is attributed to the
presence of a shape resonance in the $d$-wave collisional entrance channel
according to the prediction of Ref.\ \cite{simoni2002:pra}. Simoni {\it et
al.} \cite{simoni2002:pra} also predict a light shift enhanced by a $d$-wave
shape resonance. In both the Na and Rb case, the calculated light shift
ignoring the shape resonance is found negligible. The clear absence of a shift
in our measurements for intensities up to the saturation intensity seems to
support the absence of any $p$- or $d$-wave shape resonance in cesium.  The
calculations performed in Ref.\ \cite{pellegrini2003:phd} also predicts a
negligible light shift in the presently explored intensity range.  It should
be noted that the absence of a light shift in the present experiment also
leads to a correspondingly negligible inhomogeneous broadening of the
photoassociation resonance, in contrast to previous saturation studies where
it had to be accounted for \cite{schloder2002:pra,mckenzie2002:prl}.



Photoassociation experiments inside a quasi-electrostatic optical dipole trap
allow for precise measurements of absolute photoassociation rate coefficients,
due to the high pair density, the long interaction times, and the well defined
internal hyperfine state of the trapped atoms. This allows us in this work to
investigate quantitatively the saturation of the photoassociation rate
coefficient with increasing photoassociation laser intensity. It is found that
the scattering probability saturates at unity, corresponding to the the
unitarity limit of this inelastic scattering process. The same result was
concluded in the experiment with quantum degenerate $^7$Li
\cite{prodan2003:prl}. Saturation at the unitarity limit has to be
distinguished from saturation in an optical transition, thus Rabi oscillations
between the pair of atoms and the molecular dimer, for which we see no
indication, are not expected under these conditions. We also observe that the
photoassociation resonance width exhibits a linear increase with
intensity. Both the saturation and the width increase are in good quantitative
agreement with the theoretical model of Bohn and Julienne using the saturation
intensity as the only parameter of the model, for which a value of
450$\pm$100\,W/cm$^2$ is obtained. A light shift is found to be strongly
suppressed, which is attributed to the lack of a $p$ or $d$-wave shape
resonance in cesium, in contrast to the findings for sodium or rubidium.



We would like to thank P. Julienne and E. Tiemann for very useful
discussions. Support by the Deutsche Forschungsgemeinschaft is acknowledged
within the Schwerpunktprogramm 1116 ,,interactions in ultracold atomic and
molecular gases''.  We also acknowledge support by the EU research training
network ,,Cold molecules'' (COMOL), under the contract number HPRN-2002-00290.


\end{document}